\documentstyle[epsfig,  emulateapj5]{aastex}
\def\prd{\ref@jnl{Phys.Rev.D}}          

\def\r#1{{\it \ref{#1}}}

\def\kms{\mbox{\,km s$^{-1}$}}

\def\Mpc{\mbox{\,Mpc}}
\def\kpc{\mbox{\,kpc}}
\def\pc{\mbox{\,pc}}
\def\Ms{\mbox{\,M$_{\odot}$}}

\def\cm{\mbox{\,cm}}
\def\ccc{\mbox{\,cm$^{-3}$}}

\def\s{\mbox{\,s}}
\def\yr{\mbox{\,yr}}
\def\Myr{\mbox{\,Myr}}
\def\G{\mbox{\,G}}

\def\tento#1{\times 10^{#1}}

\def\HH{\mbox{\,H$_2$}}

\def\K{{\rm \ K}}

\def\yrs{{\rm \ years}}

\def\gtsima{$\; \buildrel > \over \sim \;$}
\def\ltsima{$\; \buildrel < \over \sim \;$}
\def\prosima{$\; \buildrel \propto \over \sim \;$}
\def\gsim{\lower.7ex\hbox{\gtsima}}
\def\lsim{\lower.7ex\hbox{\ltsima}}
\def\simgt{\lower.7ex\hbox{\gtsima}}
\def\simlt{\lower.7ex\hbox{\ltsima}}
\def\simpr{\lower.7ex\hbox{\prosima}}

\def\edcomment#1{\iffalse\marginpar{\raggedright\sl#1\/}\else\relax\fi}
\reversemarginpar

\makeatletter

\newenvironment{figurehere}
  {\def\@captype{figure}}
  {}
\makeatother

\begin{document}
\title{The Formation of the First Star in the Universe}
 \author{Tom Abel}
\affil{Harvard Smithsonian Center for Astrophysics, MA, US--02138
 Cambridge}
\affil{Institute of Astronomy, Cambridge, UK}
\author{Greg L. Bryan}
\affil{Massachusetts Institute of Technology, MA, US--02139 Cambridge}
\affil{Hubble Fellow}
\author{Michael L. Norman}
\affil{University of California, San Diego, CA, US--92093 La Jolla}

\begin{abstract}
  We describe results from a fully self--consistent three dimensional
  hydrodynamical simulation of the formation of one of the first stars in the
  Universe. Dark matter dominated pre-galactic objects form because of
  gravitational instability from small initidal density perturbations. As they
  assemble via hierarchical merging, primordial gas cools through
  ro-vibrational lines of hydrogen molecules and sinks to the center of the
  dark matter potential well.  The high redshift analog of a molecular cloud
  is formed.  When the dense, central parts of the cold gas cloud become
  self-gravitating, a dense core of $\sim 100\Ms$ undergoes rapid contraction.
  At densities $n>10^9 \ccc$ a $1\Ms$ proto-stellar core becomes fully
  molecular due to three--body \HH formation. Contrary to analytical
  expectations this process does not lead to renewed fragmentation and only
  one star is formed.  The calculation is stopped when optical depth effects
  become important, leaving the final mass of the fully formed star somewhat
  uncertain.  At this stage the protostar is acreting material very rapidly
  ($\sim 10^{-2}\Ms \yr^{-1}$).  Radiative feedback from the star will not
  only halt its growth but also inhibit the formation of other stars in the
  same pre--galactic object (at least until the first star ends its life,
  presumably as a supernova).  We conclude that at most one massive
  ($M\gg1\Ms$) metal free star forms per pre--galactic halo, consistent with
  recent abundance measurements of metal poor galactic halo stars.
\end{abstract}

\section{Motivation}

Chemical  elements  heavier  than  Lithium  are  synthesized  in  stars.  Such
``metals'' are  observed at times when  the Universe was only  $\lsim 10$\% of
its current  age in  the inter--galactic medium  (IGM) as absorption  lines in
quasar  spectra (see  Ellison et  al. 2000,  and references  therein).  Hence,
these heavy  elements not  only had  to be synthesized  but also  released and
distributed  in the IGM  within the  first billion  years. Only  supernovae of
sufficiently short lived massive stars are known to provide such an enrichment
mechanism. This leads to the prediction that

{\it the first generation of cosmic structures formed massive stars
  (although not necessarily only massive stars).}

In the past 30 years it has been argued that the first cosmological objects
form globular clusters (\r{PD68}), super--massive black holes (\r{H69}), or
even low mass stars (\r{PSS83}).  This disagreement of theoretical studies
might at first seem surprising. However, the first objects form via the
gravitational collapse of a thermally unstable reactive medium, inhibiting
conclusive analytical calculations.  The problem is particularly acute because
the evolution of all other cosmological objects (and in particular the larger
galaxies that follow) will depend on the evolution of the first stars.

Nevertheless, in comparison to present day star formation, the physics of the
formation of the first star in the universe is rather simple. In particular:
\begin{itemize}
\item the chemical and radiative of processes in the primordial gas are
readily understood. 
\item strong magnetic fields are not expected to exist at early times.
\item by definition no other stars exist to influence the
environment through radiation, winds, supernovae, etc.
\item the emerging standard model for structure formation provides
appropriate initial conditions.
\end{itemize}

In previous work we have presented three--dimensional cosmological simulations
of the formation of the first objects in the universe (\r{A95}, \r{AANZ})
including first applications of adaptive mesh refinement (AMR) cosmological
hydrodynamical simulations to first structure formation (\r{ABN99}, \r{ABN00},
ABN hereafter) .  In these studies we achieved a dynamic range of up to
$2\tento{5}$ and could follow in detail the formation of the first dense
cooling region far within a pre--galactic object that formed
self--consistently from linear density fluctuation in a cold dark matter
cosmology. Here we report results from simulations that extend our previous
work by another 5 orders of magnitude in dynamic range. For the first time it
is possible to bridge the wide range between cosmological and stellar scale.

\section{Simulation Setup and Numerical Issues}\label{sec:sim}

We employ an Eulerian structured adaptive mesh refinement cosmological
hydrodynamical code developed by Bryan and Norman (\r{BN97}, \r{BN99}). The
hydrodynamical equations are solved with the second order accurate piecewise
parabolic method (\r{WC84}; \r{Betal95}) where a Riemann solver ensures
accurate shock capturing with a minimum of numerical viscosity.  We use
initial conditions appropriate for a spatially flat Cold Dark Matter cosmology
with 6\% of the matter density contributed by baryons, zero cosmological
constant, and a Hubble constant of 50 km/s/Mpc (\r{cosmo}).  The power
spectrum of initial density fluctuations in the dark matter and the gas are
taken from the computation by the publicly available Boltzmann code CMBFAST
(\r{SZ96}) at redshift 100 (assuming an Harrison--Zel'dovich scale--invariant
initial spectrum).

We set up a three dimensional volume with 128 comoving kpc on a side and solve
the cosmological hydrodynamics equations assuming periodic boundary
conditions.  This small volume is adequate for our purpose, because we are
interested in the evolution of the first pre--galactic object within which a
star may be formed by a redshift of $z\sim 20$.  First we identify the
Lagrangian volume of the first proto--galactic halo with a mass of $\sim
10^6\Ms$ in a low resolution pure N--body simulation. Then we generate new
initial conditions with four initial static grids that cover this Langrangian
region with progressively finer resolution. With a $64^3$ top grid and a
refinement factor of 2 this specifies the initial conditions in the region of
interest equivalent to a $512^3$ uni--grid calculation.  For the adopted
cosmology this gives a mass resolution of $1.1\Ms$ for the dark matter (DM,
hereafter) and $0.07\Ms$ for the gas.  The small DM masses ensure that the
cosmological Jeans mass is resolved by at least ten thousand particles at all
times. Smaller scale structures in the dark matter will not be able to
influence the baryons because of their shallow potential wells. The
theoretical expectation holds, because the simulations of ABN which had 8
times poorer DM resolution led to identical results on large scales as the
simulation presented here.

During the evolution, refined grids are introduced with twice the spatial
resolution of the parent (coarser) grid.  These child (finer) meshes are added
whenever one of three refinement criteria are met.  Two Langrangian criteria
ensure that the grid is refined whenever the gas (DM) density exceeds 4.6
(9.2) its initial density.  Additionally, the local Jeans length is always
covered by at least 64 grid cells \footnote{The Jeans mass which is the
  relevant mass scale for collapse and fragmentation is thus resolved by at
  least $4\pi 32^3/3\approx 1.4\tento{5}$ cells.} (4 cells per Jeans length
would be sufficient, \r{TKMH97}).  We have also carried out the simulations
with identical initial conditions but varying the refinement criteria. In one
series of runs we varied the number of mesh points per Jeans length.  Runs
with 4, 16, and 64 zones per Jeans length are indistinguishable in all mass
weighted radial profiles of physical quantities.  No change in the angular
momentum profiles could be found, suggesting negligible numerical viscosity
effects on angular momentum transport.  A further refinement criterion that
ensured the local cooling time scale to be longer than the local Courant time
also gave identical results. This latter test checked that any thermally
unstable region was identified.

The simulation follows the non--equilibrium chemistry of the dominant nine
species species (H, H$^+$, H$^-$, e$^-$, He, He$^+$, He$^{++}$, H$_2$, and
H$_2^+$) in primordial gas. Furthermore, the radiative losses from atomic and
molecular line cooling, Compton cooling and heating of free electrons by the
cosmic background radiation are appropriately treated in the optically thin
limit (\r{AAZN97}, \r{AZAN97}).  To extend our previous the studies to higher
densities three essential modifications to the code were made . First we
implemented the three--body molecular hydrogen formation process in the
chemical rate equations. For temperatures below 300 K we fit to the data of
Orel (\r{Orel}) to get $k_{3b} = 1.3\tento{-32} (T/300\K)^{-0.38}\cm^6\s^{-1}$.
Above 300 K we then match it continuously to a powerlaw (\r{PSS83}) $k_{3b} =
1.3\tento{-32} (T/300\K)^{-1}\cm^6\s^{-1}$.  Secondly, we introduce a variable
adiabatic index for the gas (\r{ON98}).  The dissipative component (baryons)
may collapse to much higher densities than the collisionless component (DM).
The discrete sampling of the DM potential by particles can then become
inadequate and result in artificial heating of the baryons (cooling for the
DM) once the gas density becomes much larger than the local DM density. To
avoid this, we smooth the DM particles with a Gaussian of width 0.05 \pc\ for
grids with cells smaller than this length.  At this scale, the enclosed gas
mass substantially exceeds the enclosed DM mass.

The standard message passing library (MPI) was used to implement domain
decomposition on the individual levels of the grid hierarchy as a
parallelization strategy.  The code was run in parallel on 16 processors of
the SGI Origin2000 supercomputer at the National Center for Supercomputing
Applications at the University of Illinois at Urbana Champaign.

We stop the simulation at a time when the molecular cooling lines reach an
optical depth of ten at line center because our numerical method cannot treat
the difficult problem of time--dependent radiative line transfer in
multi--dimensions. At this time the code utilizes above 5500 grids on 27
refinement levels with $1.8\tento{7}\approx 260^3$ computational grid cells.
An average grid therefore contains $\sim 15^3$ cells.

\section{Results}

\begin{figure*}[ht]
\centerline{\psfig{file=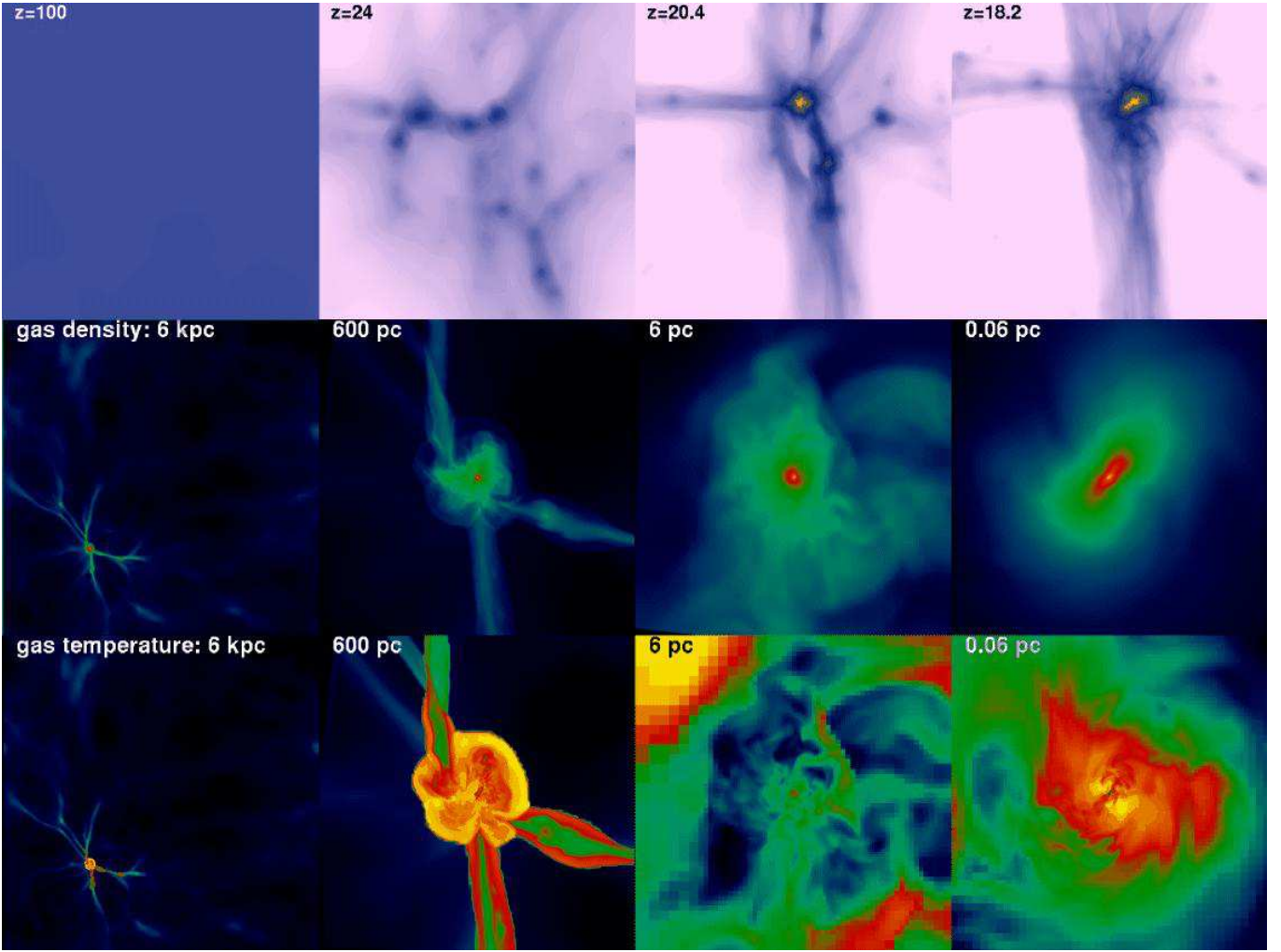,width=18cm}}
\caption{\footnotesize Overview of the evolution and collapse forming a
  primordial star in the universe. The top row shows projections of the gas
  density of one thousands of the simulation volume approximately centered at
  the pre--galactic object within which the star is formed.  The four
  projections from left to right are taken at redshifts 100, 24, 20.4, and
  18.2 respectively.  The pre--galactic objects form from very small density
  fluctuations, and continously merge to form larger objects.  The middle and
  bottom row show thin slices through the gas density and temperature at the
  final simulation output.  The leftmost panels are on the scale of the
  simulation volume $\sim 6$ proper kpc. The panels to the right zoom in
  towards the forming star and have side lengths of 600 pc, 6pc, and 0.06 pc
  (12000 astronomical units). The color maps (going from black to blue, green,
  red, yellow) are logarithmic and the associated values were adjusted
  considerably to visualize the $\sim 17$ orders of magnitude in density
  covered by these simulations.  In the left panels the larger scale
  structures of filaments and sheets are seen. At their intersections a
  pre--galactic object of $\sim 10^6\Ms$ is formed.  In the temperature slice
  (second panel - bottom row) one sees how the gas shock heats as it falls
  into the pre--galactic object.  After passing the acretion shock the
  material forms hydrogen molecules and starts to cool.  The cooling material
  accumulates at the center of the object and forms the high redshift
  molecular cloud analog (third panel from the right) which is dense and cold
  ($T\sim 200\K$). Deep within the molecular cloud a few hundred Kelvin warmer
  core of $\sim 100\Ms$ is formed (right panel) within which a $1\Ms$
  proto--star is formed (yellow region in the right panel of the middle row).
  } \vspace{.2cm} \label{colorplate}
\end{figure*}

\begin{figure*}[th]
\epsfysize=8cm
\epsfxsize=10cm
\vspace{-.3cm}
\centerline{\psfig{file=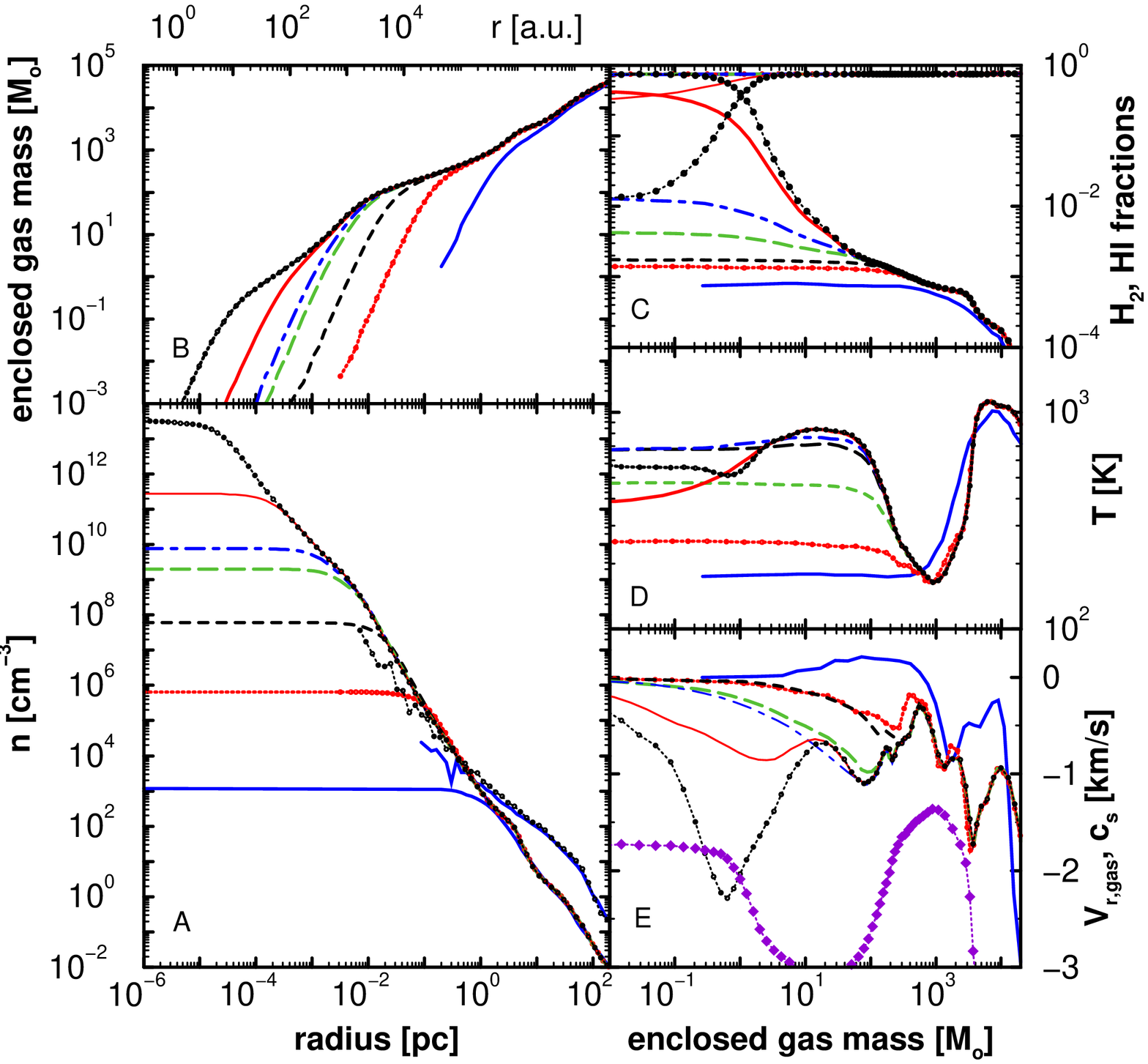,width=15cm, height=13cm}}
\caption{\footnotesize{ Radial mass weighted averages of various
    physical quantities  at seven  different output times.  Panel A  shows the
    evolution of  the particle number density  in $\cm^{-3}$ as  a function of
    radius at  redshift 19  (solid line), nine  Myrs later (dotted  lines with
    circles),  $0.3\Myr$  later  (dashed  line), $3\tento{4}\yr$  later  (long
    dashed line), $3\tento{3}\yr$  later (dot--dashed line), $1.5\tento{3}\yr$
    later (solid line) and finally  $200\yrs$ later (dotted line with circles)
    at $z=18.181164$. The two lines between $10^{-2}$ and $200\pc$ give the DM
    mass density in GeV\ccc\ at z=19 and the final time, respectively. Panel B
    gives  the enclosed  gas  mass as  a function  of  radius. In  C the  mass
    fractions of atomic hydrogen and molecular hydrogen are shown. Panel D and
    E  illustrate  the temperature  evolution  and  the  mass weighted  radial
    velocity of the baryons, respectively. The bottom line with filled symbols
    in panel  E shows the negative  value of the  local speed of sound  at the
    final time.  In all panels  the same output  times correspond to  the same
    line styles.  } }\label{5panel}
\end{figure*}

\subsection{Characteristic mass scales}

Our simulations (Fig.~\ref{colorplate}, Fig.~\ref{5panel}), identify at least
four  characterisic mass  scales.  From  the  outside going  in, one  observes
infall  and  accretion  onto the  pre--galactic  halo  with  a total  mass  of
$7\tento{5}\Ms$,  consistent with  previous  studies (\r{A95},  \r{Tegmark97},
\r{AANZ}, ABN, and \r{BCL99} for discussion and references).

At a mass scale of about 4000 solar mass ($r\sim 10\pc$) rapid cooling and
infall is observed.  This is accompanyed by the first of three valleys in the
radial velocity distribution (Fig.~\ref{5panel}E).  The temperature drops and
the molecular hydrogen fraction increases.  It is here, at number densities of
$\sim 10 \ccc$, that the high redshift analog of a molecular cloud is formed.
Although the molecular mass fraction is not even 0.1\% it is sufficient to
cool the gas rapidly down to $\sim 200\K$.  The gas cannot cool below this
temperature because of the sharp decrease in the cooling rate below $\sim
200K$.

At redshift 19 (Fig.~\ref{5panel}), there are only two mass scales; however,
as time passes the central density grows and eventually passes $10^4 \ccc$, at
which point the ro-vibrational levels of \HH\ are populated at their
equilibrium values and the cooling time becomes independent of density
(instead of inversely proportional to it).  This reduced cooling efficiency
leads to an increase in the temperature (Fig.~\ref{5panel}D).  As the
temperature rises, the cooling rate again increases (it is 1000 times higher
at 800 K than at 200 K), and the inflow velocities slowly climb.

In order to better understand what happens next, we examine the stability of
an isothermal gas sphere.  The critical mass for gravitational collapse given
an external pressure $P_{ext}$ (BE mass hereafter) is given by Ebert
(\r{Ebert}) and Bonnor (\r{Bonnor}) as:
\begin{eqnarray}
M_{BE} = 1.18\Ms \frac{c_s^4}{G^{3/2}} P_{ext}^{-1/2}; \ \ \ c_s^2
=\frac{dP}{d\rho} = \frac{\gamma k_B T}{\mu m_H}.
\end{eqnarray}
Here $P_{ext}$ is the external pressure and $G$, $k_B$, and $c_s$ the
gravitational constant, the Boltzmann constant and the sound speed,
respectively.  We can estimate this critical mass locally if we set the
external pressure to be the local pressure to find $M_{BE}\approx 20\Ms
T^{3/2}n^{-1/2}\mu^{-2}\gamma^2$ where $\mu \approx 1.22$ is the mean mass per
particle in units of the proton mass.  Using an adiabatic index $\gamma=5/3$,
we plot the ratio of the enclosed gas mass to this modified BE mass in
Figure~\ref{BEmass}.

Our modeling shows (Fig.~\ref{BEmass}), that by the fourth considered output
time, the central 100\Ms\ exceeds the BE mass at that radius, indicating
unstable collapse.  This is the third mass scale and corresponds to the second
local minimum in the radial velocity curves (Fig.~\ref{5panel}E).  The
inflow velocity is $1\kms$ is still subsonic.  Although this mass scale is
unstable, it does not represent the smallest scale of collapse in our
simulation.  This is due to the increasing molecular hydrogen fraction.

When the gas density becomes sufficiently large ($\sim 10^{10} \ccc$),
three-body molecular hydrogen formation becomes important.  This rapidly
increases the molecular fraction (Fig.~\ref{5panel}C) and hence the cooling
rate.  The increased cooling leads to lower temperatures and even stronger
inflow and.  At a mass scale of $\sim 1 \Ms$, not only is the gas nearly
completely molecular, but the radial inflow has become supersonic
(Fig.~\ref{5panel}E).  When the \HH\ mass fraction approaches unity, the
increase in the cooling rate saturates, and the gas goes through a radiative
shock.  This marks the first appearance of the proto--stellar accretion shock
at a radius of about 20 astronomical units from its center.

\subsection{Chemo--Thermal Instability}

When the cooling time becomes independent of density the classical criterion
for fragmentation $t_{cool}< t_{dyn} \propto n^{-1/2}$ (\r{Field65}) cannot be
satisfied at high densities. However, in principal the medium may still be
subject to thermal instability. The instability criterion is 
\begin{eqnarray}
\rho \left(\frac{\partial L}{\partial \rho}\right)_{T=const.} - T
\left(\frac{\partial L}{\partial T}\right)_{\rho=const.} + 
L(\rho,T) > 0, \label{eq:ic}
\end{eqnarray}
where $L$ denotes the cooling losses per second of a fluid parcel and $T$ and
$\rho$ are the gas temperature and mass density, respectively.  At densities
above the critical densities of molecular hydrogen the cooling time is
independent of density, i.e. $\partial L/\partial \rho = \Lambda(T)$ where
$\Lambda(T)$ is the high density cooling function (e.g. \r{GP98}).
Fitting the cooling function with a power-law locally around a temperature
$T_0$ so that $\Lambda(T)\propto (T/T_0)^\alpha$ one finds $\partial
L/\partial T = \rho \alpha \Lambda(T)/T$. Hence, under these circumstances the
medium is thermally stable if $\alpha > 2$.  Because, $\alpha > 4$ for the
densities and temperatures of interest, we conclude that the medium is
thermally stable.  The above analysis neglects the heating from contraction,
but this only strengthens the conclusion.  If heating balances cooling one can
neglect the $+L(\rho,T)$ term in equation~(\ref{eq:ic}) and find the medium to
be thermally stable for $\alpha > 1$.

\begin{figurehere}
\centerline{\psfig{file=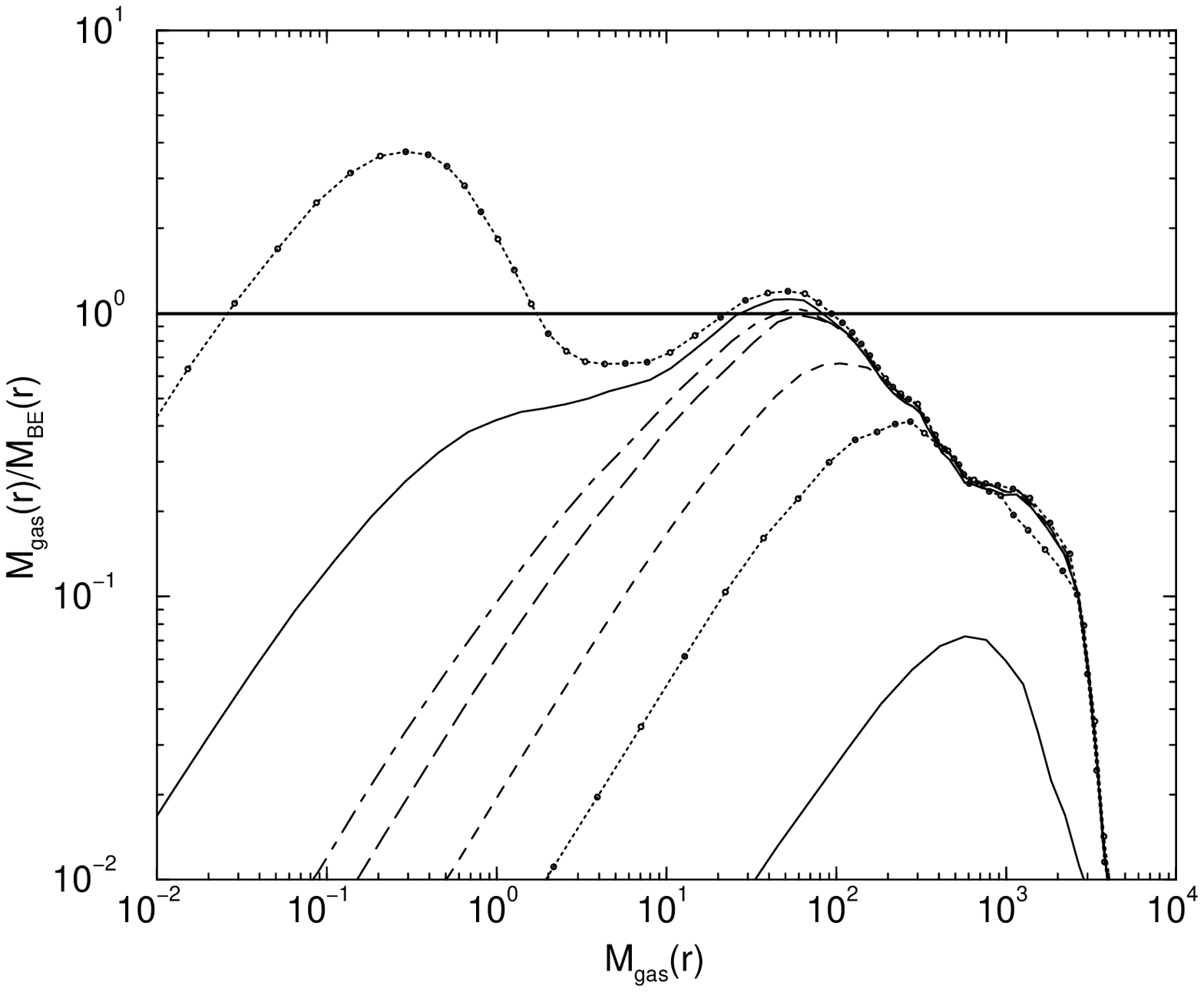,width=9cm}}
\caption{\footnotesize Ratio of enclosed gas mass to the locally
  estimated Bonnor--Ebert mass ($M_{BE}\approx 61\Ms
  T_K^{3/2}n^{-1/2}\mu^{-2}$) for various output times. The enclosed gas mass
  exceeds the BE mass at two different mass scales, $\sim 1\Ms$ and $\sim
  100\Ms$. The line-styles in the Figure correspond to the output times shown
  in Fig.~\ref{5panel}.  } \vspace{.2cm} \label{BEmass}
\end{figurehere}

However, here we neglected the chemical processes.  The detailed analysis for
the case when chemical processes occur on the collapse time--scale is well
known (\r{SY77}).  This can be applied to primordial star formation
(\r{Silk83}) including the three--body formation of molecular hydrogen
(\r{PSS83}) which drives a chemo--thermal instability. Evaluating all the
terms in this modified instability criterion (\r{Silk83}, equation~36) one
finds the simple result that for molecular mass fractions $f<6/(2\alpha + 1)$
the medium is expected to be chemo--thermally unstable. These large molecular
fractions illustrate that the strong density dependence of the three body \HH\ 
formation dominates the instability. Examining the three dimensional
temperature and \HH\ density field we clearly see this chemo--thermal
instability at work.  Cooler regions have larger \HH\ fractions. However, no
corresponding large density inhomogeneities are found and fragmentation does
not occur. This happens because of the short sound crossing times in the
collapsing core. When the \HH\ formation time--scale becomes shorter than the
cooling time the instability originates.  However, as long as the sound
crossing time is much shorter than the chemical and cooling time scales the
cooler parts are efficiently mixed with the warmer material.  This holds in
our simulation until the final output where for the first time the \HH\ 
formation time scale becomes shorter than the sound--crossing time.  However,
at this point the proto--stellar core is fully molecular and stable against
the chemo--thermal instability.  Consequently no large density contrasts are
formed. Because at these high densities the optical depth of the cooling
radiation becomes larger than unity the instability will be suppressed even
further.

\subsection{Angular momentum}
\label{sec:ang_mom}

\begin{figure*}[ht]
\epsfysize=10cm
\epsfxsize=12cm
\vspace{-.3cm}
\centerline{\psfig{file=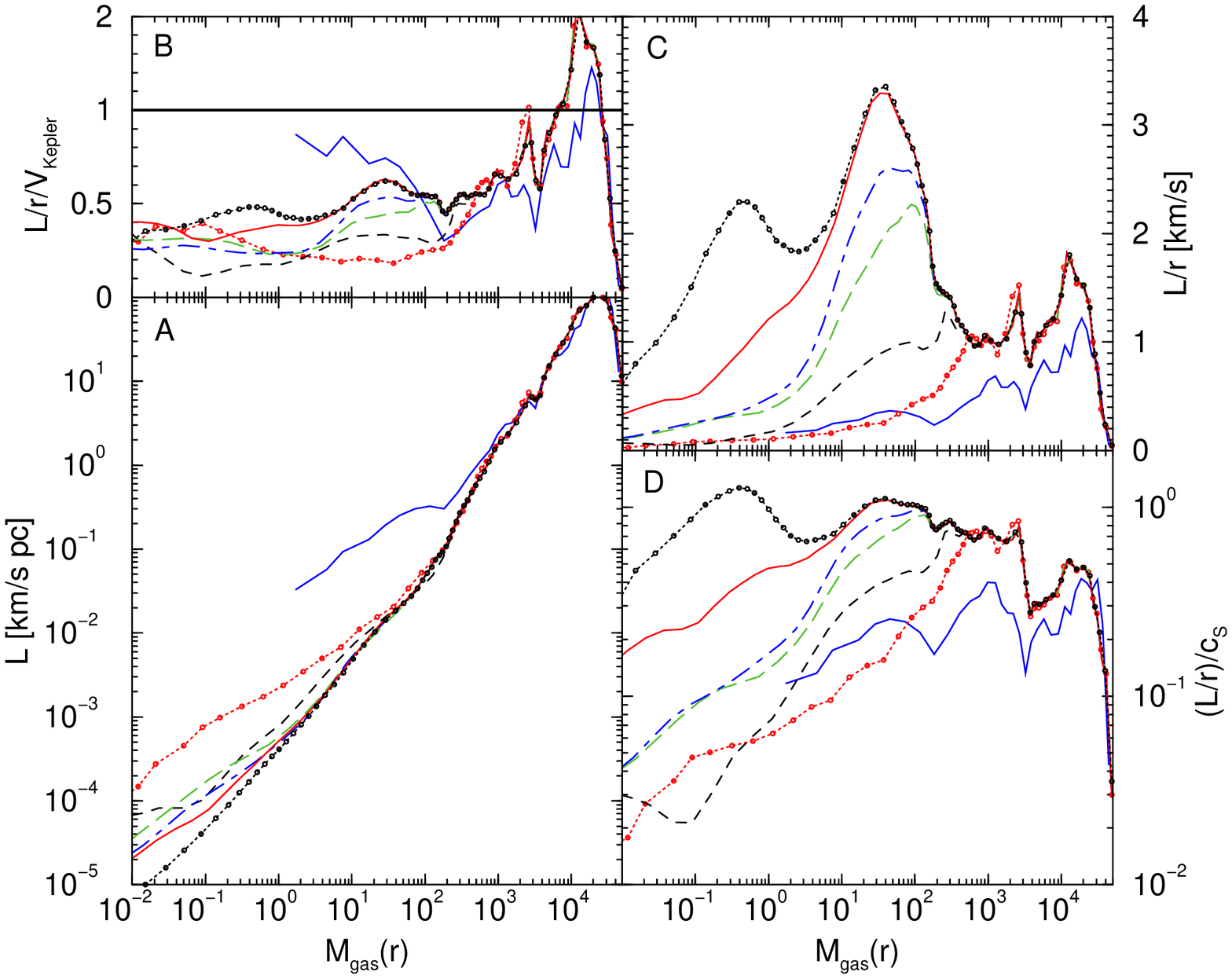,width=16cm}}
\caption{Radial mass weighted averages of various physical quantities
related to the angular momentum of the gas. The seven different output
times correspond to the ones described in Fig.~\ref{5panel}. Panel A
shows the specific angular momentum $L$ in \kms \pc\ as a function of
enclosed gas mass. The typical rotational speed $L/r$ is shown in
panel C and its ratio to the Keplerian velocity
$V_{Kepler}=(GM/r)^{1/2}$ and the local speed of sound in panel B and
D, respectively.}\label{angtrans}
\end{figure*}

Interestingly, rotational support does not halt the collapse. This is for two
reaons. The first is shown in panel A of Fig.~\ref{angtrans}, which plots
the specific angular momentum against enclosed mass for the same seven output
times discussed earlier.  Concentrating on the first output
(Fig.~\ref{angtrans}), we see that the central gas begins the collapse with a
specific angular momentum only $\sim 0.1$\% as large as the mean value.  This
type of angular momentum profile is typical of halos produced by gravitational
collapse (e.g. \r{QZ88}), and means that the protostellar gas starts out
without much angular momentum to lose.  As a graphic example of this, consider
the central one solar mass of the collapsing region.  It has only an order of
magnitude less angular momentum at densities $n\gsim 10^{13}\cm^{-3}$ than it
had at $n\gsim 10^{6}\cm^{-3}$ although it collapsed by over a factor 100 in
radius.

The remaining output times (Fig.~\ref{angtrans}) indicate that there is some
angular momentum transport within the central $100 \Ms$ (since L plotted as a
function of enclosed mass should stay constant as long as there is no shell
crossing).  In panel C, we divide $L$ by $r$ to get a typical rotational
velocity and in panels B and D compare this velocity to the Keplerian
rotational velocity and the local sound speed, respectively.

We find that the typical rotational speed is a factor two to three below that
required for rotational support. Furthermore, we see that this azimuthal speed
never significantly exceeds the sound speed, although for most the mass below
$100 \Ms$ it is comparable in value.  We interpret this as evidence that it is
shock waves during the turbulent collapse that are responsible for much of the
transported angular momentum.  A collapsing turbulent medium is different from
a disk in Keplarian rotation.  At any radius there will be both low and high
angular momentum material, and pressure forces or shock waves can redistribute
the angular momentum between fluid elements.  Lower angular momentum material
will selectively sink inwards, displacing higher angular momentum gas.  This
hydrodynamic transport of angular momentum will be suppressed in situation
where the collapse proceeds on the dynamical time rather on the longer cooling
time as in the presented case. This difference in cooling time and the widely
different initial conditions may explain why this mechanism has not been
observed in simulations of present day star formation (e.g. \r{BB93}, and
references therein). However, such situations may also arise in the late
stages of the formation of present day stars and in scenarios for the
formation of super--massive black holes.

To ensure that the angular momentum transport is not due to numerical shear
viscosity (\r{NWB80}) we have carried out the resolution study discussed
above. We have varied the effective spatial resolution by a factor 16 and
found identical results. Furthermore, we have run the adaptive mesh refinement
code with two different implementations of the hydrodynamics solver. The
resolution study and the results presented here were carried out with a direct
piecewise parabolic method adopted for cosmology (\r{WC84}; \r{Betal95}). We
ran another simulation with the lower order ZEUS hydrodynamics (\r{SN92}) and
still found no relevant differences. These tests are not strict proof that the
encountered angular momentum transport is not caused by numerical effects;
however, they are reassuring.

\subsection{Magnetic Fields?}
The strength of magnetic fields generated around the epoch of recombination is
minute. In contrast, phase transitions at the qantum--chromo--dynamic (QCD)
and electro--weak scales may form even dynamically important fields.  While
there is a plethora of such scenarios for primordial magnetic field generation
in the early universe they are not considered to be an integral part of our
standard picture of structure formation.  This is because not even the order
of these phase transitions is known (\r{SOJ97}), and references therein).
Unfortunately, strong primordial small-scale ($\ll 1$ comoving \Mpc) magnetic
fields are poorly constrained observationally (\r{BFS97}).

The critical magnetic field for support of a cloud (\r{MS76}) allows a rough
estimate up to which primordial magnetic field strengths we may expect our
simulation results to hold. For this we also assume a flux frozen flow with no
additional amplification of the magnetic field other than the contraction
($B\propto \rho^{2/3}$). For a comoving B field of $\gsim 3\tento{-11}\G$ on
scales $\lsim 100\kpc$ the critical field needed for support may be reached
during the collapse possibly modifying the mass scales found in our purely
hydrodynamic simulations. However, the ionized fraction drops rapidly during
the collapse because of the absence of cosmic rays ionizations. Consequently
ambipolar diffusion should be much more effective in the formation of the
first stars even if such strong primordial magnetic fields were present.

\section{Discussion}
Previously we discussed the formation of the pre--galactic object and the
primordial ``molecular cloud'' that hosts the formation of the first star in
the simulated patch of the universe (\r{ABN00}). These simulations had a
dynamic range of $\sim 10^5$ and identified a $\sim 100\Ms$ core within the
primordial ``molecular cloud'' undergoing renewed gravitational collapse. The
fate of this core was unclear because there was the potential caveat that
three body \HH\ formation could have caused fragmentation. Indeed this further
fragmentation had been suggest by analytic work (\r{Silk83}) and single zone
models (\r{PSS83}).  The three dimensional simulations described here were
designed to be able to test whether the three body process will lead to a
break up of the core. {\sl No fragmentation due to three body \HH\ formation
  is found.}  This is to a large part because of the slow quasi--hydrostatic
contraction found in ABN which allows sub--sonic damping of density
perturbations and yields a smooth distribution at the time when three body
\HH\ formation becomes important.  Instead of fragmentation a single fully
molecular proto--star of $\sim 1\Ms$ is formed at the center of the $\sim
100\Ms$ core.

However, even with extraordinary resolution, the {\it final} mass of the first
star remains unclear. Whether all the available cooled material of the
surroundings will accrete onto the proto--star or feedback from the forming
star will limit the further accretion and hence its own growth is difficult to
compute in detail. 

Within $10^4\yr$ about $70\Ms$ may be accreted assuming that angular momentum
will not slow the collapse (Fig.~\ref{acrete}). The maximum of the accretion
time of $\sim 5\tento{6}\yr$ is at $\sim 600 \Ms$.  However, stars larger than
$100\Ms$ will explode within $\sim 2\Myr$.  Therefore, it seems unlikely (even
in the absence of angular momentum) that there would be sufficient time to
accrete such large masses.  \vspace{.1cm}
\begin{figurehere}
\centerline{\psfig{file=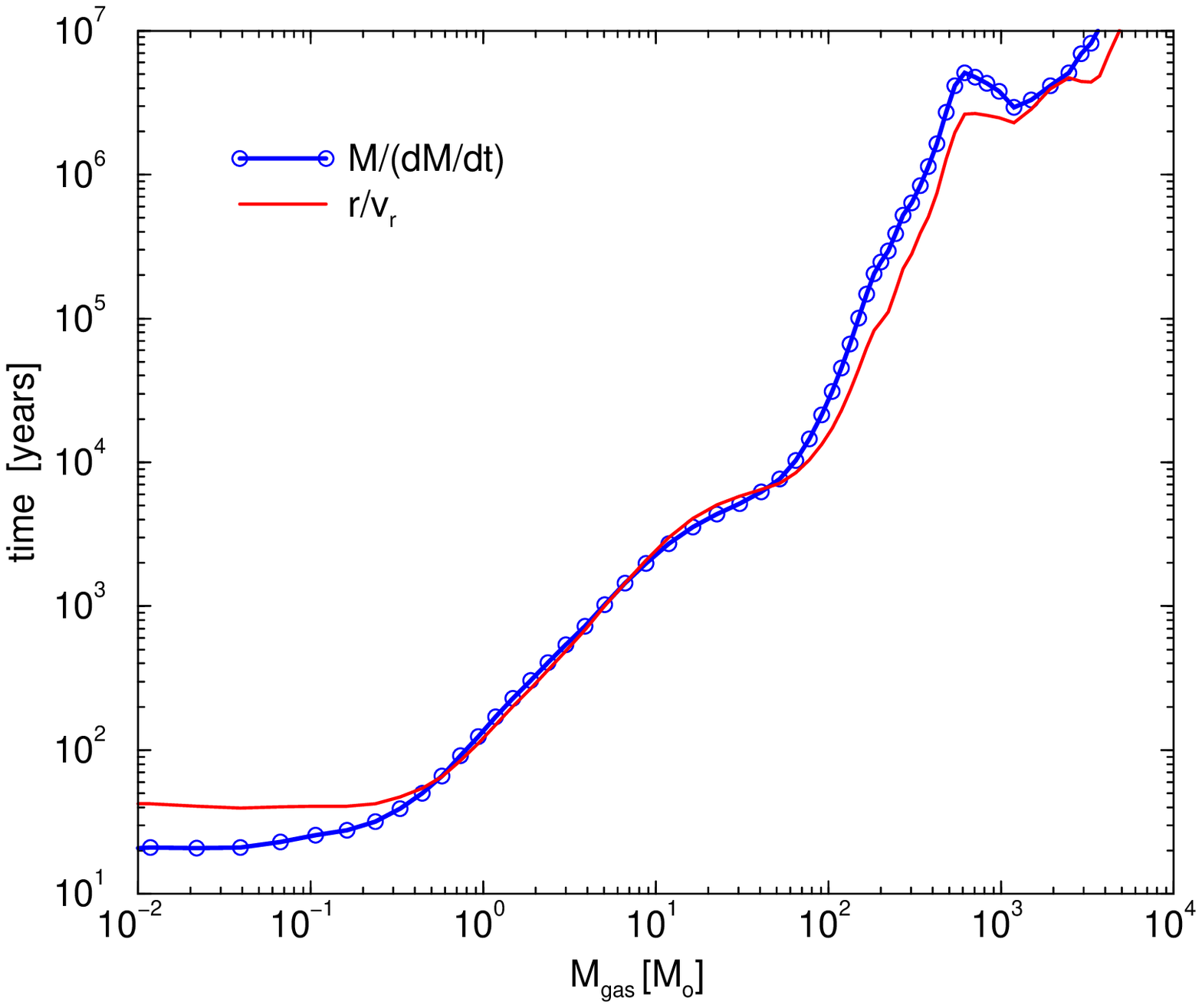,width=9cm}}
\caption{The accretion time as function of enclosed gas
  mass. The line with symbols gives $M(r)/[4 \pi \rho(r) r^2 |v_r(r)| ]$. The
  solid line simply shows how long it would take the mass to move to $r=0$ if
  it were to to keep its current radial velocity ($r/v_r(M)$). }\label{acrete}
\end{figurehere}
\vspace{.2cm}

A  one solar  mass  proto--star will  evolve  too slowly  to halt  substantial
accretion. From  the accretion time profile (Fig.~\ref{acrete})  one may argue
that a more realistic minimum mass limit of the first star should be $\gsim 30
\Ms$ because this  amount would be accreted within a  few thousand years. This
is  a very short  time compared  to expected  proto--stellar evolution  times. 
However, some properties of the primordial  gas may make it easier to halt the
accretion. One possibility is the  destruction of the cooling agent, molecular
hydrogen,  without   which  the   acreting  material  may   reach  hydrostatic
equilibrium.  This may or may not be sufficient to halt the accretion. One may
also  imagine  that  the central  material  heats  up  to $10^4$  K,  allowing
Lyman-$\alpha$ cooling  from neutral hydrogen. That cooling  region may expand
rapidly as  the internal pressure  drops because of infall,  possibly allowing
the  envelope to accrete  even without  molecular hydrogen  as cooling  agent. 
Additionally,  radiation pressure  from  ionizing photons  as  well as  atomic
hydrogen Lyman  series photons may  become significant and  eventually reverse
the flow. The mechanisms discussed  by Haehnelt (1993) on galactic scales will
play an important role for the continued accretion onto the proto--star.  This
is  an interplay  of many  complex physical  processes because  one has  a hot
ionized Str\"omgren sphere through which  cool and dense material is trying to
accrete.  In such  a situation one expects a  Raleigh--Taylor type instability
that is modified via the geometry of the radiation field.

At the final output time  presented here there are $\sim 4\tento{57}$ hydrogen
molecules in  the entire  protogalaxy. Also the  \HH\ formation time  scale is
long because  there are  no dust grains  and the  free electrons (needed  as a
catalyst) have  almost fully recombined.  Hence, as soon  as the the  first UV
photons of Lyman Werner band frequencies  are produced there will be a rapidly
expanding photo--dissociating  region (PDR) inhibiting  further cooling within
it.  This  photo-dissociation  will   prevent  further  fragmentation  at  the
molecular cloud scale.  I.e. no other  star can be formed within the same halo
before  the first  star dies  in a  supernova. The  latter, however,  may have
sufficient energy to unbind the  entire gas content of the small pre--galactic
object  it   formed  in  (\r{MF99}).   This  may  have   interesting  feedback
consequences for the dispersal of  metals, entropy and magnetic field into the
intergalactic medium (\r{F98}, \r{CB00}).

Smoothed particle hydrodynamics (SPH, e.g. \r{M92}), used extensively in
cosmological hydrodynamics, has been employed (\r{BCL99}) to follow the
collapse of solid body rotating uniform spheres.  The assumption of coherent
rotation causes these clouds to collapse into a disk which developes
filamentary structures which eventually fragment to form dense clumps of
masses between $100$ and $1000$ solar masses.  It has been argued that these
clumps will continue to accrete and merge and eventually form very massive
stars.  These SPH simulation have unrealistic initial conditions and much less
resolution then our calculations. However, they also show that many details of
the collapse forming a primordial star are determined by the properties of the
hydrogen molecule.

We have also simulated different initial density fields for a Lambda CDM
cosmology. There we have focused on halos with different clustering
environments. Although we have not followed the collapse in these halos to
proto-stellar densities, we have found no qualitative differences in the
``primordial molecular cloud'' formation process as discussed in ABN. Also
other AMR simulations (\r{MBA01}) give consistent results on scales larger
than $1\pc$. In all cases a cooling flow forms the primordial molecular cloud
at the center of the dark matter halo.  We conclude that the molecular cloud
formation process seems to be independent of the halo clustering properties
and the adopted CDM type cosmology.  Also the mass scales for the core and the
proto--star are determined by the local Bonnor--Ebert mass. Consequently, we
expect the key results discussed here to be insensitive to variations in
cosmology or halo clustering.

\section{Conclusion}

The picture arising from these numerical simulations has some very interesting
implications. It is possible that all metal free stars are massive and form in
isolation. Their  supernovae may  provide the metals  seen in even  the lowest
column density  quasar absorption lines (\r{ESSP00}, and  references therein). 
Massive primordial stars offer a natural explanation for the absence of purely
metal free low mass stars in the Milky Way. The consequences for the formation
of galaxies may  be even more profound in that  the supernovae provide metals,
entropy, and magnetic fields and may  even alter the initial power spectrum of
density fluctuations of the baryons.

Interestingly, it has  been recently argued, from abundance  patterns, that in
low metallicity  galactic halo stars  seem to have  been enriched by  only one
population of  massive stars (\r{WQ00}).   These results, if  confirmed, would
represent  strong  support  for  the   picture  arising  from  our  ab  initio
simulations of first structure formation.

To end on a speculative note there is suggestive evidence that links gamma ray
bursts to  sites of massive  star formation (e.g.  \r{R99}). It would  be very
fortunate if a  significant fraction of the massive  stars naturally formed in
the  simulations would  cause gamma  ray  bursts (e.g.  \r{CL00}).  Such  high
redshift bursts  would open a  remarkably bright window  for the study  of the
otherwise dark (faint) Ages.

\begin{enumerate}
\item \label{PD68}  Peebles, P.~J.~E.~\& Dicke, R.~H.\ 1968, ApJ {\bf 154}, 891 
\item \label{H69}   Hirasawa, T. 1969, Prog. Theor. Phys. {\bf 42}, 523
\item \label{PSS83} Palla, F., Salpeter, E.E., Stahler, S.W. 1983, ApJ {\bf 271}, 632
\item \label{A95}  Abel, T. 1995, thesis, University of Regensburg.
\item \label{AANZ} Abel, T., Anninos, P., Norman, M.L., Zhang, Y. 1998, ApJ
  {\bf 508}, 518.
\item \label{ABN99}  Abel, T., Bryan, G. L., \& Norman, M. L. 1999, in "Evolution of Large Scale Structure: From Recombination to Garching", eds. Banday, T., Sheth, R. K. and Costa, L. N.
\item \label{ABN00} ABN: Abel, T., Bryan, G.L., Norman, M.L. 2000, ApJ {\bf 540}, 39
\item \label{cosmo} Friedmann models with a cosmological constant which
  currently seem to fit various observational test better differ from the
  standard CDM model considered here only slightly at the high redshifts
  modeled. 
\item \label{BN97}  Bryan, G.L., Norman, M.L. 1997, in {\it Computational Astrophysics}, eds. D.A. Clarke and M. Fall, ASP Conference \#123
\item \label{BN99}  Bryan, G.L., Norman, M.L. 1999, in {\it Workshop on Structured Adaptive Mesh Refinement Grid Methods}, IMA Volumes in Mathematics No. 117, ed. N. Chrisochoides, p. 165
\item \label{WC84} Woodward, P. R., \& Colella, P. 1984, J. Comput. Phys. {\bf
    54}, 115
\item \label{Betal95}  Bryan, G.L., Cen, R., Norman, M.L., Ostriker, J.P.\ \&
  Stone, J.M.\ 1994, ApJ {\bf 428}, 405 
\item \label{SZ96}  Seljak, U., \& Zaldarriaga, M. 1996, ApJ {\bf 469}, 437
\item \label{TKMH97}  Truelove, J.\ K., Klein, R.\ I., McKee, C.\ F.,
  Holliman, J.\ H., Howell, L.\ H.\ \&  Greenough, J.\ A.\ 1997, ApJL {\bf 489}, L179 
\item \label{AAZN97}  Abel, T., Anninos, P., Zhang, Y.,  Norman, M.L. 1997,
  NewA {\bf 2}, 181
\item \label{AZAN97}  Anninos, P., Zhang, Y., Abel, T., and Norman, M.L. 1997,
  NewA {\bf 2}, 209.
\item \label{Orel}   Orel, A.E. 1987, J.Chem.Phys. {\bf 87}, 314
\item \label{ON98}  Omukai, K.\ \& Nishi, R.\ 1998, ApJ {\bf 508}, 141 
\item \label{Tegmark97}  Tegmark, M., Silk, J., Rees, M.J., Blanchard, A.,
  Abel, T., Palla, F. 1997,  ApJ {\bf 474}, 1.
\item \label{BCL99} Bromm, V., Coppi, P.~S., \& Larson, R.~B.\ 1999, ApJL {\bf
    527}, L5 
\item \label{Ebert} Ebert, R. 1955, Zs. Ap. 217
\item \label {Bonnor} Bonnor, W. B. 1956, MNRAS  {\bf 116}, 351
\item \label{Field65}  Field, G. B. 1965, ApJ {\bf 142}, 531
\item \label{GP98}  Galli, D.\ \& Palla, F.\ 1998, A\&A {\bf 335}, 403 
\item \label{SY77}  Sabano, Y. \& Yoshi, Y. 1977, PASJ {\bf 29}, 207
\item \label{Silk83}  Silk, J. 1983, MNRAS {\bf 205}, 705
\item \label{QZ88}  Quinn, P.J. \& Zurek, W.H. 1988, ApJ {\bf 331}, 1
\item \label{BB93}  Burkert, A. \& Bodenheimer, P. 1993, MNRAS {\bf 264}, 798 
\item \label{NWB80}  Norman, M.L., Wilson, J.R. \& Barton, R.T. 1980, ApJ {\bf
    239}, 968 
\item \label{SN92}  Stone, J.M. \& Norman, M.L. 1992, ApJS {\bf 80}, 791 
\item \label{SOJ97}  Sigl, G., Olinto, A.\ V.\ \& Jedamzik, K.\ 1997,
  Phys.Rev.D {\bf 55}, 4582 
\item \label{BFS97}  Barrow, J.\ D., Ferreira, P.\ G.\ and Silk, J.\ 1997,
  Physical Review Letters {\bf 78}, 3610 
\item \label{MS76}  Mouschovias, T. Ch., Spitzer, L. 1976, ApJ {\bf 210}, 326
\item \label{Haehnelt} Haehnelt, M.G. 1995, MNRAS {\bf 273}, 249
\item \label{MF99}  Mac Low, M. \& Ferrara, A. 1999, ApJ {\bf 513}, 142 
\item \label{F98}  Ferrara, A. 1998, ApJL {\bf 499}, L17
\item \label{CB00}  Cen, R. \& Bryan, G.L. 2000, ApJL {\bf 546}, L81
\item \label{M92}  Monaghan, J.\ J.\ 1992, ARAA {\bf 30}, 543 
\item \label{MBA01}  Machacek, M.E., Bryan, G.L., \& Abel, T 2001, ApJ {\bf 548}, 509
\item \label{ESSP00}  Ellison, S.\ L., Songaila, A., Schaye, J.\ \& Pettini,
  M.\ 2000, \aj {\bf 120}, 1175 
\item \label{WQ00}  Wasserburg, G.\ J.\ \& Qian, Y.\ 2000, ApJL {\bf 529}, L21 
\item \label{R99}  Reichart, D.\ E.\ 1999, ApJL {\bf 521}, L111 
\item \label{CL00}  Ciardi, B.\ \& Loeb, A.\ 2000, ApJ {\bf 540}, 687

\item T.A. happily acknowledges stimulating and insightful discussions with
  Martin Rees and Richard Larson.  GLB was supported through Hubble Fellowship
  grant HF-0110401-98A from the Space Telescope Science Institute, which is
  operated by the Association of Universities for Research in Astronomy, Inc.
  under NASA contract NAS5-26555.

\end{enumerate}

\end{document}